\documentclass[10pt,a4paper]{article}
\usepackage[utf8]{inputenc}

\usepackage[english]{babel}

\usepackage{multicol}

\usepackage{amsmath}
\usepackage{amsfonts}
\usepackage{amssymb}
\usepackage{amsthm}

\def\mon3sat5{{\sc Monotone 3-Sat-5}}
\def\monTsatF{{\sc Monotone 3-Sat-4}}
\def\sat34{{\sc{3-Sat-4}}}
\def\monfour{{\sc Monotone $(2,\,3)$-Sat-4}}

\newtheorem{theorem}{Theorem}
\newtheorem{lemma}{Lemma}

\begin{document}
	\title{\monTsatF\ is $\mathcal{NP}$-complete.}
	\author{Andreas Darmann, Janosch Döcker}
	\date{\today}
	\maketitle

\begin{abstract}
\monTsatF\ is a variant of the satisfiability problem for boolean formulae in conjunctive normal form. In this variant, each clause contains exactly three literals---either all or none of them are positive, i.\,e., no clause contains both a positive and a negative literal---and every variable appears at most four times in the formula. Moreover, every clause consists of three distinct literals. We show that \monTsatF\ is $\mathcal{NP}$-complete. 
\end{abstract}

\section{Introduction}

The satisfiability problem for boolean formulae in conjunctive normal form---or one of its many variants---is frequently used in order to show that some decision problem is $\mathcal{NP}$-hard; for an introduction in the theory of $\mathcal{NP}$-completeness we refer to Garey and Johnson \cite{Garey1979}. Here, the motivation for looking into monotone variants of this problem is a conjecture attributed to Sarah Eisenstat in the scribe notes~\cite{Demaine2014} of an MIT lecture\footnote{Algorithmic Lower Bounds: Fun with Hardness Proofs (Fall '14), Prof. Erik Demaine, Teaching assistants: Sarah Eisenstat, Jayson Lynch}. The conjecture states that \mon3sat5\ is $\mathcal{NP}$-hard.

The notation {\sc $r$-Sat-$s$} denotes the variant of the satisfiability problem where every clause contains exactly $r$ distinct variables and each variable appears in at most $s$ clauses. When we use $(p,\,q)$ instead of $r$ this means that every clause contains either $p$ or $q$ distinct variables. We write clauses as subsets of a finite set $\mathcal{V}$ of variables, emphasizing that all variables need to be different in the variants of the satisfiability problem we consider in this paper. A $k$-clause contains exactly $k$ distinct variables and a clause is called monotone if either all contained literals are positive or all of them are negative, respectively. A mixed clause is a clause which is not monotone, i.e., it contains at least one positive and at least one negative literal. Let $C$ be a $k$-clause. The notation $\operatorname{Var}(C)$ means that we remove negations if there are any, i.e., we map $C$ to the monotone $k$-clause containing the same variables in their unnegated form.  

{\sc Monotone $r$-Sat-$s$} is the restriction of {\sc $r$-Sat-$s$} such that all clauses are monotone. It is known that the monotone satisfiability problem for boolean formulae in conjunctive normal form is $\mathcal{NP}$-hard \cite{Gold1978} and remains hard even if every clause contains exactly three \emph{distinct} variables (see \cite{Li1997}).

In this paper, we prove the conjecture mentioned above and show that even \monTsatF\ remains hard. The latter problem is a restriction of \sat34\ which was proven to be $\mathcal{NP}$-hard by Tovey \cite{Tovey1984}. Tovey also showed that {\sc{3-Sat-3}} is trivial, i.\,e., instances of this problem are always satisfiable. Consequently, {\sc Monotone 3-Sat-3} is trivial as well. 
 
\section{Hardness of {\sc Monotone 3-Sat-$s$} for $s \geq 4$}

Let $\mathcal{I} := (\mathcal{V},\, \mathcal{C})$ be any \sat34\ instance (for the proof that \sat34\ is $\mathcal{NP}$-complete see the work by Tovey \cite{Tovey1984}). Applying Gold's \cite[p.\,314f]{Gold1978} replacement rule to each mixed clause yields an equisatifiable \monfour\ instance $\mathcal{I}' :=  (\mathcal{V}',\, \mathcal{C}')$: Consider any mixed clause $C = C^+ \cup C^-$, where $C^+$ contains the positive literals of $C$ and $C^-$ the negative literals, respectively. Then, creating a new variable $u$ and replacing $C$ with the two clauses $C^+ \cup \{u\}$ and $C^- \cup \{\bar{u}\}$ yields an equisatisfiable instance with one mixed clause less. Since $|C^+| + |C^-| = |C| = 3$, one of the introduced clauses has size 2 and the other one has size 3. The replacement does not change the number of appearances of any variable $v \in \mathcal{V}$ and the created variable appears exactly twice in $\mathcal{I}'$. Thus, we have shown:

\begin{lemma}
\monfour\ is $\mathcal{NP}$-complete.
\end{lemma}

The next step is to replace the clauses of size 2. Li \cite[p.\,295]{Li1997} observed that a clause $\{x,\,y\}$ is satisfiable if and only if
\[
\{x,\,y,\,u\},\,\{x,\,y,\,v\},\,\{x,\,y,\,w\},\,\{\bar{u},\,\bar{v},\,\bar{w}\}
\]
are satisfiable and $\{\bar{x},\,\bar{y}\}$ is satisfiable if and only if
\[
\{\bar{x},\,\bar{y},\,\bar{u}\},\,\{\bar{x},\,\bar{y},\,\bar{v}\},\,\{\bar{x},\,\bar{y},\,\bar{w}\},\,\{u,\,v,\,w\}
\]
are satisfiable, where $u$, $v$ and $w$ are distinct new variables. Note that this replacement rule increases the number of appearances of the variables $x$ and $y$. We show that this can be avoided by defining a suitable replacement rule which only creates new variables with at most five appearances. 

In the following we define multiple rules $\mathcal{R}_i$ that replace a monotone 2-clause $C$ in a collection $\mathcal{K}$ of clauses by monotone 3-clauses $C_1,\,C_2,\,\ldots,C_j$ so that $C$ is satisfiable if and only if $C_1,\,C_2,\,\ldots,C_j$ are satisfiable and 
\[
\left (\bigcup_{k = 1}^{j}\operatorname{Var}(C_k)\right ) \cap \left ( \bigcup_{C_{\ell}\in \mathcal{K}}\operatorname{Var}(C_{\ell}) \right ) \subseteq \operatorname{Var}(C),
\]
i.e., with the exception of the two variables appearing in $C$ all other variables appearing in $C_k$, $1\leq k\leq j$, are new variables. The rules are of the form
\[
   \mathcal{R}_i^{\mathcal{K}} :=
   \begin{cases}
     \{x,\,y\} \equiv C_1,\,C_2,\,\ldots,C_{j_i} \\
     \{\bar{x},\,\bar{y}\} \equiv C_1',\,C_2',\,\ldots,C_{j_i}'
   \end{cases},
\]
where $\mathcal{K}$ is a collection of clauses, i.e., the context in which the rule is applied. In the following we omit the $\mathcal{K}$ in the rule definitions to increase readability. Note that applying such a rule changes the context for further applications of the same or different rules. The notation $\{x,\,y\} \equiv C_1,\,C_2,\,\ldots,C_{j_i}$ means that the clause $\{x,\,y\}$ is satisfiable if and only if the clauses $C_1,\,C_2,\,\ldots,C_{j_i}$ are satisfiable; the other case is defined in the same way. We write $\mathcal{R}_i(C)$ to denote a rule application respecting the properties mentioned above: If $C$ consists of two positive literals, then we replace $C$ according to the top case of the rule; and if $C$ consists of two negative literals we replace $C$ according to the bottom case. We use the notation $\Delta_{\mathcal{R}_i}^x$, $\Delta_{\mathcal{R}_i}^y$ and $\Delta_{\mathcal{R}_i}^{\text{new}}$  to denote the maximum number by which an application of rule $\mathcal{R}_i$ to a clause  $\{x,\,y\}$ or $\{\bar{x},\,\bar{y}\}$ increases the appearances of $x$, $y$, and the new variables, respectively.   

\paragraph*{Replacement rule $\mathcal{R}_1$}

Let $\mathcal{R}_1$ denote Li's replacement rule, which looks in our notation as follows:
\[
   \mathcal{R}_1 :=
   \begin{cases}
     \{x,\,y\} \equiv \{x,\,y,\,u\},\,\{x,\,y,\,v\},\,\{x,\,y,\,w\},\,\{\bar{u},\,\bar{v},\,\bar{w}\} \\
     \{\bar{x},\,\bar{y}\} \equiv \{\bar{x},\,\bar{y},\,\bar{u}\},\,\{\bar{x},\,\bar{y},\,\bar{v}\},\,\{\bar{x},\,\bar{y},\,\bar{w}\},\,\{u,\,v,\,w\}.
   \end{cases}
\]
We have
\[
\Delta_{\mathcal{R}_1}^x = \Delta_{\mathcal{R}_1}^y = \Delta_{\mathcal{R}_1}^{\text{new}} = 2.
\]

\paragraph*{Replacement rule $\mathcal{R}_2$}

As an intermediate step we define a second replacement rule:
\[
   \mathcal{R}_2 :=
   \begin{cases}
     \{x,\,y\} \equiv \{x,\,y,\,u\},\,\{x,\,y,\,v\},\,\mathcal{R}_1(\{\bar{u},\,\bar{v}\}) \\
     \{\bar{x},\,\bar{y}\} \equiv \{\bar{x},\,\bar{y},\,\bar{u}\},\,\{\bar{x},\,\bar{y},\,\bar{v}\},\,\mathcal{R}_1(\{u,\,v\}).
   \end{cases}
\]
Observe that 
\begin{align*}
\{x,\,y\} \text{ is satisfiable } \Leftrightarrow 
\{x,\,y,\,u\},\,\{x,\,y,\,v\},\,\mathcal{R}_1(\{\bar{u},\,\bar{v}\})
\text{ are satisfiable }
\end{align*}
and 
\begin{align*}
\{\bar{x},\,\bar{y}\} \text{ is satisfiable } \Leftrightarrow 
\{\bar{x},\,\bar{y},\,\bar{u}\},\,\{\bar{x},\,\bar{y},\,\bar{v}\},\,\mathcal{R}_1(\{u,\,v\})
\text{ are satisfiable, }
\end{align*}
where $u$ and $v$ are distinct new variables. We have  
\[
\Delta_{\mathcal{R}_2}^x = \Delta_{\mathcal{R}_2}^y = 1 \text{ and } \Delta_{\mathcal{R}_2}^{\text{new}} = \max (2 + \Delta_{\mathcal{R}_1}^u,\, 2 + \Delta_{\mathcal{R}_1}^v,\, \Delta_{\mathcal{R}_1}^{\text{new}}) = 4.
\]

\paragraph*{Replacement rule $\mathcal{R}_3$} 

Using the preceding rule---and implicitly also Li's rule---we can define a replacement rule with the desired properties:
\[
   \mathcal{R}_3 :=
   \begin{cases}
     \{x,\,y\} \equiv \{x,\,y,\,u\},\,\mathcal{R}_2(\{\bar{u},\,\bar{v}\}),\,\mathcal{R}_2(\{\bar{u},\,\bar{w}\}),\,\mathcal{R}_2(\{v,\,w\}) \\
     \{\bar{x},\,\bar{y}\} \equiv \{\bar{x},\,\bar{y},\,\bar{u}\},\,\mathcal{R}_2(\{u,\,v\}),\,\mathcal{R}_2(\{u,\,w\}),\,\mathcal{R}_2(\{\bar{v},\,\bar{w}\}).
   \end{cases}
\]
Observe that 
\begin{align*}
\{x,\,y\} \text{ is satisfiable } &\Leftrightarrow 
\{x,\,y,u\},\,\{\bar{u}\}
\text{ are satisfiable } \\
&\hspace*{-2cm}\Leftrightarrow 
\{x,\,y,\,u\},\,\mathcal{R}_2(\{\bar{u},\,\bar{v}\}),\,\mathcal{R}_2(\{\bar{u},\,\bar{w}\}),\,\mathcal{R}_2(\{v,\,w\})
\text{ are satisfiable}
\end{align*}
and 
\begin{align*}
\{\bar{x},\,\bar{y}\} \text{ is satisfiable }\\
&\hspace*{-2cm}\Leftrightarrow 
\{\bar{x},\,\bar{y},\,\bar{u}\},\,\mathcal{R}_2(\{u,\,v\}),\,\mathcal{R}_2(\{u,\,w\}),\,\mathcal{R}_2(\{\bar{v},\,\bar{w}\}
\text{ are satisfiable,}
\end{align*}
where $u$, $v$ and $w$ are distinct new variables.
We have 
\begin{align*}
\Delta_{\mathcal{R}_3}^x = \Delta_{\mathcal{R}_3}^y = 0 \text{ and } \Delta_{\mathcal{R}_3}^{\text{new}} &= \max (3+2\Delta_{\mathcal{R}_2}^u,\, 2 + 2\Delta_{\mathcal{R}_2}^v,\, 2+ 2\Delta_{\mathcal{R}_2}^w,\,\Delta_{\mathcal{R}_2}^{\text{new}}) = 5. 
\end{align*}

An application of Rule $\mathcal{R}_3$ replaces one clause with 19 new clauses using 18 new variables and reduces the number of 2-clauses by one. Actually, 17 clauses and 16 variables suffice, since we could have used $\mathcal{R}_1(C)$ instead of $\mathcal{R}_2(C)$ for $C\in \{\{v,\,w\},\,\{\bar{v},\,\bar{w}\}\}$ in the definition of $\mathcal{R}_3$. The reason for not doing so is that $\mathcal{R}_3$ and the calculation of $\Delta_{\mathcal{R}_3}^{\text{new}}$ appear a little simpler the way it is now. The number of necessary applications of $\mathcal{R}_3$ is exactly the number of 2-clauses (of a monotone instance, of course). Since applying $\mathcal{R}_3$ only introduces variables appearing at most five times and leaves the number of appearances of all other variables unchanged, we have proven:

\begin{theorem}\label{the:mon3sat5}
\mon3sat5\ is $\mathcal{NP}$-complete.
\end{theorem}

Now, we show that \monTsatF\ is $\mathcal{NP}$-complete. Again, we start with an instance of \monfour\ and the goal is to get rid of the clauses of size 2 while preserving equisatisfiability. In order to achieve that, we present a finite collection of monotone 3-clauses $\mathcal{C}_z$ such that no variable appears more than four times and a designated variable $z$ appears exactly three times, and show that this collection is satisfiable if and only if $z$ is set to true. If there is a clause of the form $\{\bar{x},\,\bar{y}\}$ in the instance, we replace this clause with $\{\bar{x},\,\bar{y},\,\bar{z}\}$ and add $\mathcal{C}_z$ to the instance. The result is an equisatisfiable \monfour\ instance with one negative 2-clause less. Of course, all variables appearing in $\mathcal{C}_z$ are newly created. By negating every variable appearance in $\mathcal{C}_z$, we can force $z$ to be set to false. Therefore, we can get rid of clauses of the form $\{x,\,y\}$ analogously. The collection $\mathcal{C}_z$ is given by the following $25$ clauses. 

\begin{multicols}{4}
\begin{enumerate}
\item $\{u,\,w,\,z\}$
\item $\{u,v,\,z\}$
\item $\{\bar{w},\,\bar{v},\,\bar{g}\}$
\item $\{\bar{w},\,\bar{v},\,\bar{h}\}$
\item $\{\bar{w},\,\bar{v},\,\bar{i}\}$
\item $\{g,\,h,\,i\}$
\item $\{\bar{m},\,\bar{n},\,\bar{g}\}$
\item $\{\bar{m},\,\bar{n},\,\bar{h}\}$
\item $\{\bar{m},\,\bar{n},\,\bar{i}\}$
\item $\{m,\,a,\,b\}$
\item $\{n,\,a,\,b\}$
\item $\{\bar{u},\,\bar{a},\,\bar{r}\}$
\item $\{\bar{u},\,\bar{b},\,\bar{r}\}$
\item $\{r,\,z,\,f\}$
\item $\{\bar{d},\,\bar{e},\,\bar{a}\}$
\item $\{\bar{d},\,\bar{e},\,\bar{b}\}$
\item $\{p,\,q,\,d\}$
\item $\{p,\,q,\,e\}$
\item $\{\bar{f},\,\bar{p},\,\bar{c}\}$
\item $\{\bar{f},\,\bar{q},\,\bar{c}\}$
\item $\{r,\,c,\,j\}$
\item $\{\bar{j},\,\bar{p},\,\bar{k}\}$
\item $\{\bar{j},\,\bar{q},\,\bar{k}\}$
\item $\{k,\,c,\,\ell\}$
\item $\{\bar{\ell},\,\bar{j},\,\bar{f}\}$
\end{enumerate}
\end{multicols}

Assume that the above collection of clauses is satisfiable by a truth assignment
in which $z$ is set false. \\
First, we show that this implies that $u$ has to be set true. If
$u$ is set false, then the first two clauses imply that both $w$
and $v$ need to be set true. Clauses $3,4,5$ thus yield that all
three of $g,h,i$ have to be set false, in contradiction with clause
$6$. Thus, $u$ has to be set true. \\
By clause $6$ at least one of $g,h,i$ has to be set true. Thus,
clauses $7,8,9$ imply that at least one of $m,n$ has to be set false.
As a consequence, clauses $10,11$ yield that at least one of $a,b$
needs to be set true. In turn, by clauses $12,13$ this means that
$r$ has to be set false (recall that $u$ is set true). Since both
$r,z$ are set false, $f$ must be set true due to clause $14$. By
the fact that at least one of $a,b$ is true, clauses $15,16$ imply
that at least one of $d,e$ is set false. In turn, by the next two
clauses this means that at least one of $p,q$ must be set true. In
addition, recalling that $f$ is set true, clauses $19,20$ imply that
$c$ has to be set false. Also recalling that $r$ is set false, this
means that $j$ has to be set true due to clause $21$. Now, clauses
$22,23$ imply---since at least one of $p,q$ is true---that $k$
has to be set false. Hence, as a consequence of clause $24$ and the
fact that both $k,c$ are set false, $\ell$ has to be set true. That
is, all of $\ell,j,f$ are set true, in contradiction with clause
$25$. Therewith, there in no satisfying truth assignment for the
above formula in which $z$ is set false. 

On the other hand, it is not hard to verify that the formula is satisfiable;
e.g., setting all variables of the set $\{z,g,a,r,e,p,k\}$ true and
the remaining ones false yields a satisfying truth assignment. 

Finally, note that $z$ occurs exactly $3$ times, while none of the
other variables is contained in more than four clauses. Thus, we have shown: 

\begin{theorem}\label{the:mon3sat4}
\monTsatF\ is $\mathcal{NP}$-complete.
\end{theorem}

\section{Conclusion}

We have proven that \monTsatF\ is $\mathcal{NP}$-complete. The correctness of the conjecture mentioned in the introduction stating that \mon3sat5\ is $\mathcal{NP}$-hard follows immediately from this result. Nonetheless, we also provided a  proof of the conjecture since the proof is interesting in itself.

\section*{Acknowledgement}

We would like to thank Britta Dorn for the valuable discussions and suggestions.

\bibliographystyle{alpha}
\bibliography{mylit}

\end{document}